\pdfoutput=1
\documentclass[
times,
aps,
twocolumn,
showpacs,
superscriptaddress,
groupedaddress,
amsmath,
amssymb,
floatfix]{revtex4-1}

\usepackage{amsthm}
\usepackage{graphicx}
\usepackage{bm}
\usepackage[T1]{fontenc}
\usepackage[utf8]{inputenc}
\usepackage{mathrsfs}
\usepackage[breaklinks=true]{hyperref}
\usepackage{url}
\usepackage[dvipsnames]{xcolor}
\definecolor{C1}{RGB}{52, 89, 149}
\definecolor{C2}{RGB}{251, 77, 61}
\definecolor{C3}{RGB}{3, 206, 164}
\definecolor{C4}{RGB}{202, 21, 81}
\hypersetup{colorlinks=true, linkcolor=C2, citecolor=C2, urlcolor=C2}
\usepackage{dsfont}
\usepackage{epstopdf}
\usepackage{tikz}
\usepackage[caption=false]{subfig}

\usepackage{fancyhdr}

\newcommand*{\mc}{\mathcal}
\newcommand*{\dg}{\dagger}
\newcommand*{\ex}{\mathrm{e}}
\newcommand*{\mf}{\mathfrak}
\newcommand*{\mbb}{\mathds}

\DeclareMathOperator{\tr}{tr}

\DeclareMathOperator{\FS}{\mf{S}}
\DeclareMathOperator*\bigcircop{\bigcirc}

\DeclareMathAlphabet\mathbfcal{OMS}{cmsy}{b}{n}

\usepackage{accents}

\newtheorem*{theorem}{Theorem}

\begin{document}
\title[]{Equilibration on average in quantum processes with finite temporal resolution} 

\author{Pedro Figueroa--Romero}
\email[]{pedro.figueroaromero@monash.edu}
\author{Kavan Modi}
\author{Felix A. Pollock}

\affiliation{School of Physics \& Astronomy, Monash University, Victoria 3800, Australia}

\date{\today}

\begin{abstract}
We characterize the conditions under which a multi-time quantum process with a finite temporal resolution can be approximately described by an equilibrium one. By providing a generalization of the notion of equilibration on average, where a system remains closed to a fixed equilibrium for most times, to one which can be operationally assessed at multiple times, we place an upper-bound on a new observable distinguishability measure comparing a multi-time process with a finite temporal resolution against a fixed equilibrium one. While the same conditions on single-time equilibration, such as a large occupation of energy levels in the initial state remain necessary, we obtain genuine multi-time contributions depending on the temporal resolution of the process and the amount of disturbance of the observer's operations on it.
\end{abstract}


\maketitle

A fundamental question at the core of statistical mechanics is that of how equilibrium arises from purely quantum mechanical laws in closed systems. This phenomenon is generically known as \emph{equilibration} or \emph{thermalization}, where in the latter case the system relaxes to a thermal state. The dissipative nature of equilibration, however, is at odds with the unitary nature of quantum mechanics. There are three main approaches to resolving this conundrum: \emph{typicality}~\cite{Popescu2006,GogolinPureQStat, gemmer2009, Goldstein_2010, Gogolin, Garnerone_2013}, which argues that small subsystems of a composite are in thermal equilibrium for almost all pure states of the whole; \emph{dynamical equilibration on average}~\cite{Tasaki_1998, Reimann_2008, PhysRevE.79.061103, ShortSystemsAndSub, ShortFinite, XXEisert}, which demonstrates that time-dependent quantities of quantum systems evolve towards fixed values and stay close to them for most times, even if they eventually deviate greatly from it; and the \emph{eigenstate thermalization hypothesis}~\cite{Deutsch1991,Srednicki, Srednicki_1999,PhysRevE.87.012118,Rigol2008, Turner2018}, which argues that the expectation values of a `physical observable' at long times are indistinguishable for an isolated system from a thermal one.

What these approaches have in common, is that they look at the statistical properties of the state of the system at long times; however, finding a system in or close to an equilibrium state does not necessarily imply all observable properties of the system have equilibrated. In particular, when measurements are coarse (i.e. only a subpart is measured), it may be that temporal correlations due to a sequence of observations may contain signatures indicating whether the system is in equilibrium or not. The recent studies of out-of-time-order correlation functions use multi-time statistics to distinguish between thermalised systems and coherent complex systems~\cite{Kitaev}. However, it may be that these multi-time statistics also equilibrate in general; that is, they are most often found close to some average value.\textsuperscript{\footnote{The out-of-time-order correlations require propagating the system back and forth in time. Here we only go forward in time.}}

In this manuscript, we focus on the case of finite temporal resolution for the dynamical equilibration of quantum processes where multiple operations are applied in sequence. We present sufficient conditions for general multi-time observations to relax close to their equilibrium values when the corresponding operations are implemented with an imperfect, or \emph{fuzzy}, clock (or, equivalently, on a system with uniformly fluctuating energies). In particular, we place an upper bound on how distinguishable the statistics of such observations are from those made in equilibrium.

\begin{figure}[t]
\centering
\includegraphics[width=0.485\textwidth]{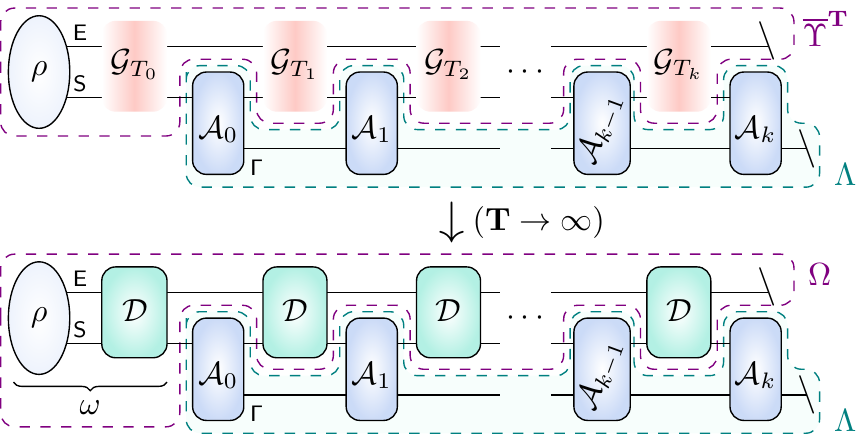}
\caption{We characterize the attainability of quantum process equilibration, i.e., under what conditions a $k$-step process $\overline{\Upsilon}^{\bf{T}}$ time-averaged over each Hamiltonian evolution $(\mc{G})$ within time-windows of width $\mathbf{T}=(T_0,T_1,\ldots,T_k)$ remains close to one dephased $(\mc{D})$ with respect to the corresponding Hamiltonian at each time step, according to a set of operations $\{\mc{A}_i\}$ on a subsystem $\mathsf{S}$, which can be correlated in time through an ancillary space $\mathsf{\Gamma}$ and represented by a single tensor $\Lambda$. By definition, equality is attained in the limit of all $\mathbf{T}\to\infty$.}
\label{Fig: processes}
\end{figure}

We first briefly recapitulate the landmark results of Refs.~\cite{ShortFinite,Reimann_2008}, in which the attainability of observable equilibration on average for a single observation is characterized mainly by two factors: the energy eigenstates of the system having a large overlap with the initial state, and the \emph{scale} set by the operator norm of the observable being measured. Building on the works of Refs.~\cite{ShortFinite,Reimann_2008}, we first ask, for the case of a single measurement, how different is an equilibrium process from an out-of-equilibrium one, when the available clock is fuzzy. We then generalise to considering observations over multiple points in time; here, temporal correlations become relevant. While the original conditions for equilibration to occur still hold, we obtain additional conditions related to the temporal resolution of the observations and how much these disturb the intermediate states. Our results hold for Hamiltonian dynamics of the total system, while the measurements are allowed to be general quantum operations that are coarse and may only act on a sub-part of the whole system.

\section{Equilibration on average}\label{sec: equilibration on avg}
In the past decade, the program of equilibration has focused on upper bounding fluctuations of observable expectation values around equilibrium, from which conclusions about equilibration of the state of the system itself have been drawn~\cite{ShortSystemsAndSub, Gogolin}. The basic mechanism behind equilibration is that of dephasing~\cite{Reimann_2008, Oliveira_2018}, and equilibration will occur as long as the initial state, following a perturbation, has an overlap with many energy eigenstates of the Hamiltonian driving the dynamics. The only further assumption is that there are not too many degenerate energy gaps~\cite{PhysRevA.98.022135}, ensuring that the majority of the system plays a dynamical role~\cite{PhysRevE.79.061103}. Specifically, observable equilibration in the sense of Ref.~\cite{ShortFinite} considers time-independent Hamiltonian dynamics given by a unitary operator $U=\exp\{-i H t\}$, with results depending on energetic properties of the Hamiltonian $H$, such as the number of distinct energy levels $\mathfrak{D}\leq{d}$ and the maximum number of energy gaps $N(\epsilon)$ in an energy window of width $\epsilon>0$.

While the choice of an equilibrium state is arbitrary, an intuitive candidate is the infinite-time-averaged subsystem state 
\begin{gather}
\omega := \lim_{T\to\infty}\overline{\rho}^T \qquad \mbox{with} \qquad \overline{X}^T :=\frac{1}{T}\int_0^T X(t)\,dt
\end{gather}
where $\rho$ stands for the initial state of the whole system, time-averaged over a finite time window of width $T$. Notably, this corresponds to a dephasing of the initial state with respect to $H$, i.e. $\omega=\mc{D}(\rho)$ where we define 
\begin{gather}
\mc{D}(\cdot):=\sum_nP_n(\cdot)P_n,
\end{gather}
with $P_n$ a projector onto the $n$th eigenspace of $H=\sum_n E_n P_n$. In fact, we can similarly describe general finite-time-averaged evolution by the map  
\begin{gather}
\begin{split}
&\mc{G}_T(\cdot):=\sum_{n,m}G_{nm}^{(T)}P_n(\cdot)P_m \quad \mbox{with} \\ &G_{nm}^{(T)}:=\overline{\exp[it(E_m-E_n)]}^T
\end{split}\label{eq: G maps}
\end{gather}
Whenever it is clear that the averaging window is $T$, we will simply denote these by $\mc{G}$ and $G_{nm}$.

In this minimal setting, the authors of Ref.~\cite{ShortFinite} prove an important result on observable equilibration by showing the closeness between the evolved state $\rho(t)$ and $\omega$. Specifically, they upper-bound the temporal fluctuations of the expectation value of a general operator $A$ around equilibrium within a finite-time window as 
\begin{gather}
\overline{|\tr[A(\rho(t) - \omega)]|^2}^T \leq \frac{\|A\|^2 N(\epsilon)f(\epsilon{T})}{d_\text{eff}(\rho)},    
\end{gather}
where $f(\epsilon{T}) = 1+8\log_2 (\mathfrak{D})/\epsilon{T}$ and $\|\cdot\|$ denotes largest singular value; crucially, the so-called inverse effective dimension (or inverse participation ratio) of the initial state, $d_\text{eff}^{-1}(\rho) := \sum_n[\tr(P_n\rho)]^2$, quantifies the number of energy levels contributing significantly to the dynamics of the initial state $\rho$.

In general, we have the hierarchy $1\leq{d}_\text{eff}(\rho)\leq\mathfrak{D}\leq{d}$, and this result implies that equilibration is attained for large $d_\text{eff}$. It has been argued, on physical grounds, that the effective dimension takes a large value in realistic situations~\cite{Gogolin,Reimann_2008}, increasing exponentially in the number of constituents of generic many-body systems~\cite{XXEisert}, and it has been proven that it takes a large value for local Hamiltonian systems whenever correlations in the initial state decay rapidly~\cite{PhysRevLett.118.140601}. The temporal fluctuations of the expectation values of $A$ around equilibrium constitute a meaningful quantifier of equilibration: a small variance relates to the expectation value of $A$ concentrating around its mean~\textsuperscript{\footnote{Strictly, the full statistics should then display equilibration.}}.

This behaviour of long-time fluctuations around equilibrium has been studied both analytically and numerically in various physical models~\cite{PhysRevLett.109.247205,PhysRevE.87.012106, PhysRevE.88.032913, PhysRevE.89.022101, PhysRevB.101.174312}, as well as for the more restrictive case of thermalization~\cite{Rigol2008,Mori_2018,Gluza_2019,PhysRevLett.123.200604}. Similarly, related questions such as an absence of equilibration~\cite{PhysRevB.76.052203, Nandkishore, Hess, PhysRevLett.120.080603}, or the robustness of equilibration and further relaxation after a perturbation have been investigated~\cite{Robinson1973,PhysRevLett.118.130601,PhysRevLett.118.140601, PhysRevA.98.022135, PhysRevE.98.062103}. Here, we take a related approach towards investigating the behaviour of quantum quantum process when the interrogations are fuzzy in time. Doing so gives focus on an operationally meaningful scenario where we show that observations that have a finite temporal resolution make it hard to tell an out-of-equilibrium process from one that is in equilibrium.

\section{Equilibration due to finite temporal resolution}
Motivated by the above result, we consider the operationally relevant implications of limited resolution in time. Firstly, we focus on the dynamics of a $d_S$-dimensional subpart $\mathsf{S}$ of a $d_Ed_S$-dimensional system $\mathsf{SE}$, and we refer to subsystem equilibration as the relaxation of $\mathsf{S}$ towards some steady state, while the whole $\mathsf{SE}$ evolves unitarily with a general time-independent Hamiltonian $H$; our results can then naturally reduce to closed-system equilibration if coarse operations on the whole $\mathsf{SE}$ are considered. We then ask how different an evolving quantum state appears from equilibrium when measured at a time that can vary in each realisation, being randomly drawn from some distribution that quantifies the \emph{fuzziness} associated with finite temporal resolution.

Specifically, by a finite-temporal resolution observation we mean an observable $A$ (either on subsystem $\mathsf{S}$ or acting coarsely on $\mathsf{SE}$) measured after a time $t>0$ sampled from a probability distribution with density function $\mathscr{P}_{T}$, i.e. which is such that $\int_0^\infty\,dt\,\mathscr{P}_T(t)=1$. Here the parameter $T$ represents the \emph{uncertainty} or \emph{fuzziness} of the distribution; for example, it could be associated with the variance of the distribution. With this definition, we may generalize the time-average over a time-window $T$ by 
\begin{gather}
\overline{X}^{\mathscr{P}_T} := \int_{0}^{\infty}dt\,\mathscr{P}_T(t)\,X.    
\end{gather}
In particular, we require that the distributions $\mathscr{P}_T$ are such that the finite-time averaging map gives the dephasing map in the infinite-time limit, $\lim_{T\to\infty}\mc{G}=\mc{D}$, or equivalently, such that $\lim_{T\to\infty}G_{nm}=\delta_{nm}$; this also renders the equilibrium state $\omega=\lim_{T\to\infty}\overline{\rho(t)}^T$ to be independent of the particular choice of distribution.

The average distinguishability by means of an observable $A$ between the equilibrium and non-equilibrium cases can be quantified as $|\tr[A(\overline{\rho}^{\mathscr{P}_T}-\omega)]|$. This can be upper-bounded by
\begin{gather}
    \left|\left<A\right>_{\overline{\rho}^{\mathscr{P}_T}-\omega}\right| \leq \mathscr{S}\, \|A\|\|\rho-\omega\|_2,
    \label{eq: main standard case}
\end{gather}
where $\left<X\right>_\sigma := \tr[X\sigma]$ and $\mathscr{S}:=\max_{n\neq{m}}|G_{nm}|$. Here $\|\sigma\|_2=\sqrt{\tr(\sigma\sigma^\dg)}$ and $\|\rho-\omega\|_2^2\leq1-(d_Ed_S)^{-1}$ is the difference in purity of the full state $\rho$ with respect to that of the equilibrium $\omega$.

\noindent\textit{Proof.} Given that $\left|\left<A\right>_{\overline{\rho}^{\mathscr{P}_T}-\omega}\right|=|\tr[A(\mc{G}-\mc{D})(\rho)]|$ and $\tr[X\sigma]\leq\|X\|\|\sigma\|_2$, Eq.~\eqref{eq: main standard case} follows because
\begin{align}
    \left\| \left(\mc{G} - \mc{D}\right)(\rho)\right\|_2^2 =&\tr\left|\sum_{n \neq m} {G}_{nm} P_n\rho P_m\right|^2\nonumber\\
    =& \sum_{\substack{n \neq m \\ n^\prime \neq m^\prime}} G_{nm}G_{m^\prime{n}^\prime}\tr\left[ P_n\rho P_mP_{m^\prime}\rho P_{n^\prime}\right]\nonumber\\
    =&\sum_{n \neq m } |G_{nm}|^2\tr\left[ P_n \rho P_m\rho\right]\nonumber\\
    \leq& \max_{n\neq{m}}|G_{nm}|^2\sum_{n \neq m}\tr[P_n\rho P_m \rho]\nonumber\\
    =&\max_{n\neq{m}}|G_{nm}|^2 \ \tr(\rho^2-\omega^2)\nonumber\\
    =&\|\rho-\omega\|_2^2 \ \max_{n\neq{m}}|G_{nm}|^2, \label{eq:singlestep}
\end{align}
where we used $\tr(\rho^2-\omega^2) = \|\rho-\omega\|_2^2$, as $\tr(\rho\,\omega) = \tr(\omega^2)$.

In general $\|\omega\|_2^2\leq{d}_\text{eff}^{-1}(\rho)$, with equality both for pure $\rho$ or when the Hamiltonian is non-degenerate; both quantities relate to how spread the initial state $\rho$ is in the energy eigenbasis.

In particular, when the fuzziness $T$ corresponds to that of the uniform distribution over an interval $[0,T]$, the probability density function is $\mathscr{P}_T=T^{-1}$, as in the results of Ref.~\cite{ShortFinite} outlined above, and we get $|G_{nm}|=|\mathrm{sin}(T\mc{E}_{nm})/T\mc{E}_{nm}|$ with $\mc{E}_{nm} := (E_n-E_m)/2$. The bound in Eq.~\eqref{eq: main standard case} then tells us that the evolved state $\rho(t)$ will differ from the equilibrium $\omega$ when measured at a given time with a temporal-resolution $T$ at most with proportion $|T\mc{E}_{nm}|^{-1}$ for the smallest energy gap $\mc{E}_{nm}$, with a scale set by the size of the observable $A$ and how different the initial state $\rho$ is from the equilibrium $\omega$.

One, however, might not stop at a single observation but continue gathering data to assess how close the system remains to equilibrium with respect to a set of possible operations, $\{\mc{A}_i\}$, as we suggestively depict in Fig.~\ref{Fig: processes}. The reason for fuzziness in the initial time is that we do not know when the process actually began. However, one question we can ask is whether, by making a sequence of measurements, we are able to overcome the fuzziness of the initial interval. These operations can correspond to any possible experimental intervention, which can be correlated with any other interventions previously made, through an ancillary system. In this case the information between time-steps propagated through the environment and the disturbance introduced by the operations might become relevant. However, the subsequent measurements will also suffer from some level of fuzziness and this must be accounted for. We now precisely establish the description for multi-time quantum processes in such generality, followed by a generalisation of Eq.~\eqref{eq: main standard case} through an upper bound to the distinguishability between a finite-time resolution process and an equilibrium one.

\section{Multi-time quantum processes}
Consider an initial state $\rho$ of the joint $\mathsf{SE}$ system unitarily evolving through a time-independent Hamiltonian dynamics until, at time $t_0$, an operation $\mc{A}_0$ is made on $\mathsf{S}$ along with an ancilla $\mathsf{\Gamma}$, which is initially uncorrelated in state $\gamma$. We denote the full initial state by $\varrho:=\rho\otimes\gamma$. After the first operation, the environment and system evolve unitarily again for a time $t_1$ until another operation $\mc{A}_1$ is made on $\mathsf{S\Gamma}$, and so on for $k$ time-steps. The joint expectation value of the series of operations is given by 
\begin{gather}
    \langle{\mc{A}_k,\ldots,\mc{A}_0}\rangle := \tr[\mc{A}_k\,\mc{U}_k \cdots \mc{A}_0\,\mc{U}_0\,(\varrho)],
    \label{eq: exp val}
\end{gather}
where $\mc{U}_\ell(\cdot)=\ex^{-iH_\ell{t}_\ell}(\cdot)\,\ex^{iH_\ell{t}_\ell}$ acts on $\mathsf{SE}$, while by an operation we explicitly mean $\mathcal{A}_\ell(\cdot) := \sum_\mu a_{\ell_\mu} K_{\ell_\mu}(\cdot)K_{\ell_\mu}^\dg$, with $\sum_\mu K_{\ell_\mu}^\dg\,K_{\ell_\mu}\leq \mbb{1}$, which acts solely on $\mathsf{S\Gamma}$; here $K_{\ell_\mu}$ are Kraus operators, potentially corresponding to measurement outcomes, and $a_{\ell_\mu}$ are the corresponding outcome weights. The Hamiltonians $H_\ell$ are in general different at each step. The ancillary space $\mathsf{\Gamma}$ can be interpreted as a quantum memory device, and might carry information about previous interactions with the system.

The information about the intrinsic dynamical process, i.e., the initial $\mathsf{SE}$ state $\rho$ and the joint unitary evolutions $\mc{U}_i$ with their respective timescales at each step, can be encoded in a positive semi-definite tensor $\Upsilon$, and similarly, the sequence of operations $\{\mc{A}_i\}$ can be encoded in a tensor of the form $\Lambda$, as depicted in Figure~\ref{Fig: processes} and detailed in Appendix~\ref{appendix: process tensor}. This simplifies the joint expectation value in Eq.~\eqref{eq: exp val} as the inner product
\begin{equation}
    \langle\Lambda\rangle_\Upsilon := \tr[\Lambda\Upsilon] = \tr[\mc{A}_k\,\mc{U}_k \cdots \mc{A}_0\,\mc{U}_0\,(\varrho)]
\end{equation}
which can be seen as a generalisation of the Born rule to multi-time step quantum processes~\cite{Costa}. Here, $\Upsilon$ becomes an unnormalized many-body density operator, and $\Lambda$ an observable. Temporal correlations, or memory, in operations are carried through space $\mathsf{\Gamma}$; any $\Lambda$ can be represented as a sequence of uncorrelated operations on a joint $\mathsf{S\Gamma}$ system. Both classically correlated operations, where the measurement basis is conditioned on past outcomes, and coherent quantum correlated measurements can be represented in this way~\cite{PhysRevA.99.042108}. The case of infinite memory and the case of completely uncorrelated operations are then extreme limits of this general setting. Formally, $\Upsilon$ is the Choi state~\cite{watrous_2018} of a quantum process, containing all its accessible dynamical information~\cite{ Markovorder1, PhysRevA.99.042108} and is the quantum generalisation of a stochastic process~\cite{Quolmogorov,processtensor, processtensor2, OperationalQDynamics}.

We finally notice that when a process ends at the first intervention $\mc{A}_0$ we have $\Upsilon(t_0) =\rho_S(t_0)$, becoming the corresponding quantum state, so that all of the previous results (single measurement) apply for such case.

\begin{figure}[t]
\centering
    \includegraphics[width=0.495\textwidth]{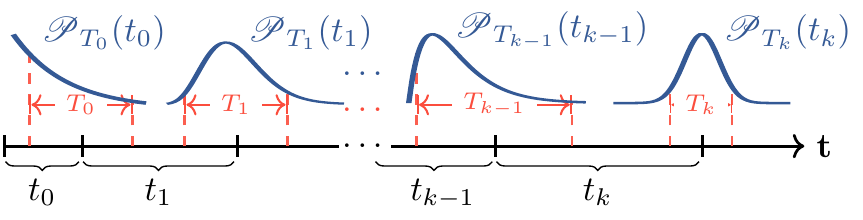}
    \caption{We consider equilibration for quantum processes with a fuzzy clock, by which we mean that each Hamiltonian evolution is time-averaged ($\mc{G}_{T_i}$) over the waiting times between interventions $t_0,\,t_1,\cdots,\,t_k$ with corresponding average waiting times $\tau_0,\,\tau_1,\cdots,\,\tau_k$ over a probability distribution with density function $\mathscr{P}_{T_i}$ with $T_i$ having a suitable uncertainty parameter role. For simplicity, in the main text we present the case $T_i=T_j$ and $\mathscr{P}_{T_i}=\mathscr{P}_{T_j}$, and denote $\mc{G}=\mc{G}_T$.}
    \label{fig: fuzziness}
\end{figure}

\section{Equilibration of multi-time observables}
Consider a quantum process as above with a fixed single Hamiltonian for all time intervals such that $H_i=H=\sum_nE_nP_n$. In Appendix~\ref{Appendix: bound} we present the general case where we do not make this assumption. This multi-time process consists of free evolution sandwiched by generalised measurements. We denote the time intervals of the free evolutions as $t_j$, which is preceded by the $j$-th measurement and followed by $j+1$-th measurement. In other words, $t_j$ is the waiting time between $j$-th and $j+1$-th measurements. As before, each of these waiting times is allowed to be fuzzy, taking a value $t_i>0$ sampled from probability distribution $\mathscr{P}_{T_i}(t_i)$, i.e. $\int_0^\infty dt_i \mathscr{P}_{T_i}(t_i)=1$. We denote the average length as 
\begin{gather}
\tau_i := \int_{0}^{\infty} dt_i \ t_i \ \mathscr{P}_{T_i}(t_i).    
\end{gather}
We pictorially represent this in Fig.~\ref{fig: fuzziness}. We denote multi-time probability distribution as $\mathscr{P}_\mathbf{T}(\mathbf{t})=\prod_{i=0}^k\mathscr{P}_{T_i}(t_i)$, where 
\begin{gather}
\mathbf{t} := (t_0, t_1, \dots, t_k) \quad \mbox{and} \quad
\mathbf{T} = (T_0,T_1,\ldots,T_k)
\end{gather}
are waiting times and the fuzziness parameters for each time interval, respectively; in what follows, we take $T_i=T_j = T$~$\forall i,j$ for simplicity (see Appendix~\ref{Appendix: bound} for the general case). The corresponding finite-temporal-resolution process is then given by
\begin{align}
\overline{\Upsilon}^{\mathscr{P}_\mathbf{T}}:= \int_0^\infty dt_k \cdots 
\int_0^\infty dt_1
\int_0^\infty dt_0
\,\mathscr{P}_\mathbf{T}(\mathbf{t})\,\Upsilon.
\end{align}
We are interested in quantifying how different this out-of-equilibrium process, where time intervals are fuzzy, looks from an equilibrium process. To define the equilibrium process we follow the lead of earlier results, i.e., the initial state relaxes to the equilibrium state $\varpi: = \mc{D}(\varrho) = \omega\otimes\gamma$ until an operation $\mc{A}_0$ is made, and subsequently the system relaxes again to an equilibrium state $\varpi_1:=\mc{D}\mc{A}_0(\varpi)$ until an operation $\mc{A}_1$ is made, and so on for $k$-time-steps. This then leads to the definition
\begin{equation}
    \varpi_i:=\mc{D}\mc{A}_{i-1}\cdots\mc{A}_0\mc{D}(\varrho),\quad\text{for}\quad i=0,\cdots,k
\end{equation}
as the equilibrium states after each intervention up to $\mc{A}_{i-1}$. This is a sensible definition for the intermediate equilibrium states, which, however, is dependent on each operation $\mc{A}_j$. We can now define the equilibrium quantum process as
\begin{gather}
    \Omega:=\lim_{\mathbf{T}\to\infty} \overline{\Upsilon}^{\mathscr{P}_\mathbf{T}}.
    \label{eq: def Omega}
\end{gather}
This is depicted in Fig.~\ref{Fig: processes} as a set of dephasing maps $\mc{D}$ at each timestep. This means then that we can write $\langle\Lambda\rangle_\Omega=\tr[\Lambda\Omega]=\tr[\mc{A}_k\varpi_k]$ for the expectation of a sequence of operations $\{\mc{A}_i\}$ on the equilibrium process $\Omega$, where equivalently, $\varpi_i=\lim_{T_i\to\infty}\overline{\mc{A}_{i-1}(\varpi_{i-1})}^{T_i}$. Since we can also express each finite averaging in the energy eigenbasis using the partial dephasing map $\mc{G}$, defined in Eq.~\eqref{eq: G maps}, we can similarly write $\langle\Lambda\rangle_{\overline{\Upsilon}^{\mathscr{P}_\mathbf{T}}}=\tr[\mc{A}_k\varrho_k]$, where we now define
\begin{gather}
\varrho_i := \mc{G}\mc{A}_{i-1} \cdots \mc{A}_0 \mc{G}(\varrho),\quad\text{for}\quad i=0,\cdots,k
\end{gather}
as intermediate, finite-time-averaged states after each intervention up to $\mc{A}_{i-1}$. As by definition $\lim_{T\to\infty}\mc{G}=\mc{D}$, the infinite-time limits $\mathbf{T}\to\infty$ make $\Upsilon$ indistinguishable from $\Omega$. We also depict this in Fig.~\ref{Fig: processes}.

We may then generalize the left hand side in Eq.~\eqref{eq: main standard case} to general quantum processes with 
$|\langle\Lambda\rangle_{\overline{\Upsilon}^{\mathbf{T}}-\Omega}|$, asking how different the statistics of a set of operations $\{\mc{A}_i\}$ can be on a fuzzy clock process, $\overline{\Upsilon}^{\mathbf{T}}$, as opposed to those in the equilibrium one $\Omega$. We provide one such answer with the following,
\begin{theorem}\label{Thm: main multitime equilibration}
    Given an environment-system-ancilla $\mathsf{(SE\Gamma)}$ with initial state $\varrho=\rho\otimes\gamma$ and initial equilibrium state $\varpi=\omega\otimes\gamma$, for any $k$-step process $\Upsilon$ with an evolution generated by a time-independent Hamiltonian on $\mathsf{SE}$, and for any fuzzy multi-time observable $\Lambda$ corresponding to a sequence of temporally local operations $\{\mc{A}_i\}_{i=0}^k$ each with fuzziness $T$ acting on a joint $\mathsf{S\Gamma}$ system,
\begin{align}
&    \left|\langle\Lambda\rangle_{\overline{\Upsilon}^{\mathscr{P}_\mathbf{T}}-\Omega}\right|\leq \mbb{A}_k + \sum_{\ell=0}^{k-1}\|\mc{A}_{k:\ell+1}\|\left(\mbb{B}_\ell + \mbb{C}_\ell\right)
,
\label{eq: result main}\\
& \mbox{with} \qquad
    \mbb{A}_k:=\mathscr{S}^{k+1}\|\mc{A}_{k:0}\|\,\|\varrho-\varpi\|_2,
    \label{eq: bound term A}
\end{align}
where here $\mathscr{S}:=\max_{n\neq{m}}|G_{nm}|$ and $\mc{A}_{j:i}:=\mc{A}_j\cdots\mc{A}_i$ is a composition of operations; the norm $\|\cdot\|$ here stands for the norm on superoperators induced by the Frobenius norm, $\|\mc{X}\|=\sup_{\|\sigma\|_2=1}\|\mc{X}(\sigma)\|_2$; the first is a single-time equilibration contribution, whereas the second term contains $k$ multi-time contributions where
\begin{align}
    \mbb{B}_\ell &:= \|[\mc{G}^{k-\ell}-\mc{D},\mc{A}_\ell]\varrho_\ell\|_2,
    \label{eq: bound t1}, \\
    \mbb{C}_\ell &:= \|[\mc{D},\mc{A}_\ell](\varrho_\ell-\varpi_\ell)\|_2,
    \label{eq: bound t2}
\end{align}
with $\varrho_i=\mc{G}\mc{A}_{i-1}\cdots\mc{A}_0\mc{G}(\varrho)$ and $\varpi_i=\mc{D}\mc{A}_{i-1}\cdots\mc{A}_0\mc{D}(\varrho)$ intermediate finite-time averaged and equilibrium states at step $i$ and where $[\cdot,\cdot]$ denotes a commutator of superoperators. 
\end{theorem}

The proof is given in full in Appendix~\ref{Appendix: bound}.

In general, by definition the term $\mathscr{S}$, which depends on the waiting time distribution $\mathscr{P}_T$, converges to zero in increasing $T$, with the rate of convergence depending on the specific distribution. In particular, for the uniform distribution on all time-steps, as in the approach by~\cite{ShortFinite}, we average over a time-window of width $T$ around each $\tau_i$ for all time-steps, with $\mathscr{P}_T=T^{-1}$ in the interval $[\tau_i-T/2, \tau_i+T/2]$, and $\mathscr{P}_T=0$ outside it. This yields $|G_{mn}|=|\mathrm{sin}(T\mc{E}_{mn})/T\mc{E}_{mn}|$; the term $\mathscr{S}$ then picks the smallest non-zero energy gap in the Hamiltonian. Similarly, if the fuzziness corresponds to that of a half-normal distribution with variance $T$, then overall $\mathscr{S}$ decays exponentially with
\begin{gather}
|G_{mn}|\sim\exp(-T\mc{E}_{mn}^2)|1-\mathrm{erf}(i\sqrt{T}\mc{E}_{mn})|,    
\end{gather}
where $\mathrm{erf}$ is the error function and $E_m-E_n=2\mc{E}_{mn}$. For both cases, if $T$ is small, $\mathscr{S}$ will also be vanishingly small whenever the energy gap $\mc{E}_{nm}$ that maximizes $|G_{nm}|$ is large enough, i.e. $\mc{E}_{nm}\gg{T}$. This property holds in general, since distributions $\mathscr{P}_T$ can be approximated as uniform for small $T$ or because the gaps $\mc{E}_{nm}$ can be seen as a rescaling factor on $T$ in the definition of $G_{nm}$.

The term $\mbb{A}_k$ in Eq.~\eqref{eq: bound term A} neglects temporal correlations and the operations $\{\mc{A}_i\}$ are all composed as a single operation $\mc{A}_{k:0}=\mc{A}_k\cdots\mc{A}_0$. The two-norm distance satisfies $\|\varrho-\varpi\|_2^2\leq{1-(d_Ed_S)^{-1}}$ as the ancillary input $\gamma$ can be taken to be pure. As discussed above, this term is suppressed through the $\mathscr{S}$ contributions when $i)$ the averaging window, or equivalently the fuzziness of the clock $T$ is large enough and $ii)$ for small $T$ whenever the energy gap maximizing the time-averaging $|G_{nm}|$ factor is large with respect to $T$.

The second term in Eq.~\eqref{eq: result main} contains genuine multi-time contributions to the bound for equilibration, which we bound further in Appendix~\ref{Appendix: bound}. These terms relate to how well intermediate states, at step $\ell$, equilibrate. Crucially, we show that the term in Eq.~\eqref{eq: bound t1} can be upper-bounded as $\mbb{B}_\ell\lesssim\mathscr{S}^{k-\ell}$, so that it is suppressed overall in the width of the time-window $T$. For the term in Eq.~\eqref{eq: bound t2}, notice that expanding and using the triangle inequality, 
\begin{gather}
\begin{split}
\mbb{C}_\ell \leq& \|\mc{D}(\varrho_{\ell+1})\|_2 + \|\varpi_{\ell+1}\|_2 \\
&+ \|\mc{A}_\ell\| (\|\mc{D}(\varrho_\ell)\|_2 + \|\varpi_{\ell}\|_2),
\end{split}
\end{gather}
where each term is the purity of a dephased state, which will decay as the effective dimension of that state.\textsuperscript{\footnote{ In general, $\tr[(\mc{D}(\sigma))^2]\leq{d}_\text{eff}^{-1}(\sigma)$ for any state $\sigma$, with equality for either pure states or non-degenerate Hamiltonians.}} On the other hand, when the control operations from $0$ to $\ell$ succeed in driving $\varrho_\ell$ so that the action of the commutator on it does not dephase it so much, the purity of $\varrho_\ell$ may be large and thus $\mbb{C}_\ell$ may become trivial (i.e. it approaches 1).

More concretely, the operations $\mc{A}_j$ interleaved within the intermediate states $\varrho_\ell$ and $\varpi_\ell$ will relate in Eq.~\eqref{eq: bound t1} and Eq.~\eqref{eq: bound t2} to how greatly they disturb either the finite-time averaged $\varrho_{j-1}$ or the equilibrated $\varpi_{j-1}$. This is most evident in the terms $\mbb{C}_\ell$, which can be bounded as $\mbb{C}_\ell\leq \|[\mc{D}, \mc{A}_\ell]\| \|\varrho_\ell - \varpi_\ell\|_2$. The norm of the commutator can be written in terms of both the capacity of the operations $\mc{A}_\ell$ to generate coherences between different energy eigenspaces from equilibrium and the degree to which the operations can turn such coherences into populations. Environments in physical systems are typically much larger than the subsystems that can be probed, and, keeping in mind that the operations $\mc{A}_j$ act only on subsystem $\mathsf{S}$ and the ancilla $\mathsf{\Gamma}$, the ability to generate and detect energy coherences should be severely limited in many physically relevant cases.

\section{Conclusions}
We have introduced an extended notion of equilibration that pertains to observations made across multiple times with a finite-temporal resolution. In a similar way to the standard case of equilibration for observables at a single time, we have put bounds on the degree to which it holds that depend on the Hamiltonian driving the evolution. We have shown that either subsystems or global coarse properties of a closed time-independent Hamiltonian system will display equilibration for multiple sequential operations with a temporal uncertainty or fuzziness provided $i)$ both the initial and intermediate states have a significant overlap with the energy eigenstates, $ii)$ the temporal fuzziness is large enough relative to the average measurement time or, equivalently, the energy gaps in the Hamiltonian are large enough with respect to small temporal fuzziness, and $iii)$ the disturbance by the operations on intermediate states is small.

Our approach for the operations that can act on the process is general in the sense that these are completely positive maps which can be correlated between time-steps and propagate information from their interactions with the subsystem $\mathsf{S}$ through the ancillary space $\mathsf{\Gamma}$. While these set a scale in all terms of the right-hand side of Eq.~\eqref{eq: result main}, they can also contribute to loosen it, potentially allowing to distinguish the fuzzy process from the equilibrium one within a finite time. It is not entirely clear, however, if a departure from equilibration is more readily accessible with a larger ancillary space $\mathsf{\Gamma}$ and for long time fuzziness $T$ the upper-bound in Eq.~\eqref{eq: result main} should remain close to zero.

Similar to the single-time standard case, equilibration over multiple observations is expected intuitively through decoherence arguments~\cite{Yukalov_2012}. The interplay with memory effects, through both the environment and the ancillary space in the interventions is as yet not entirely clear, e.g., under which circumstances finite-temporal resolution equilibration can occur without the dynamics being Markovian, i.e., memoryless, or if the temporal correlations among interventions through the ancillary space can display a departure from equilibration within a finite-time. We have previously shown rigorously that most processes in large dimensional environments are close to Markovian, and hence strongly equilibrate, in the strong coupling limit in a full typicality sense~\cite{AlmostMarkov}, as well as on complex systems obeying a large deviation bound~\cite{Markovianization}, but outside this regime the relationship between the two properties is less transparent.\\

\begin{acknowledgments}
We are grateful to Daniel Burgarth and Lucas C\'{e}leri for valuable discussions. PFR is supported by the Monash Graduate Scholarship (MGS) and the Monash International Postgraduate Research Scholarship (MIPRS). KM is supported through Australian Research Council Future Fellowship FT160100073.
\end{acknowledgments}

\onecolumngrid
\appendix
\renewcommand{\thesubsection}{\Roman{subsection}}

\section{The process tensor}\label{appendix: process tensor}
The process tensor is defined as a linear, completely positive (CP) and trace non-increasing map $\mc{T}$ from a set of CP maps $\{\mc{A}_i\}$ referred to as control operations, e.g. measurements, to a quantum state, and its action can be described as a multi-time open system evolution, e.g. for joint unitary evolution of an environment $\mathsf{E}$ plus system $\mathsf{S}$, with $\dim(\mc{H}_E\otimes\mc{H}_S)=d_Ed_S$, a $k$-step process is determined by
\begin{gather}
	\mc{T}_{k:0}[\{\mc{A}_i\}_{i=0}^{k-1}]=\tr_E[\,\mc{U}_k\mc{A}_{k-1}\cdots\mc{A}_0\,\mc{U}_0(\rho)]
\end{gather}
where $\rho$ is an initial joint $\mathsf{SE}$ state, $\mc{U}$ are unitary maps acting on $\mathsf{SE}$, and the maps $\mc{A}$ act solely on subsystem $\mathsf{S}$. We employ weighted operations, i.e. such that $\mc{A}(\cdot)=\sum{a}_\mu\,K_\mu(\cdot)K_\mu^\dg$ where $K_\mu$ are the Kraus operators of $\mc{A}$ satisfying $\sum_\mu{K}_\mu^\dg{K}_\mu\leq\mbb1$ and $a_\mu\in\mathbb{R}$ are the outcome weights for $\mc{A}$.

\begin{figure}[ht]
\includegraphics[width=0.45\textwidth]{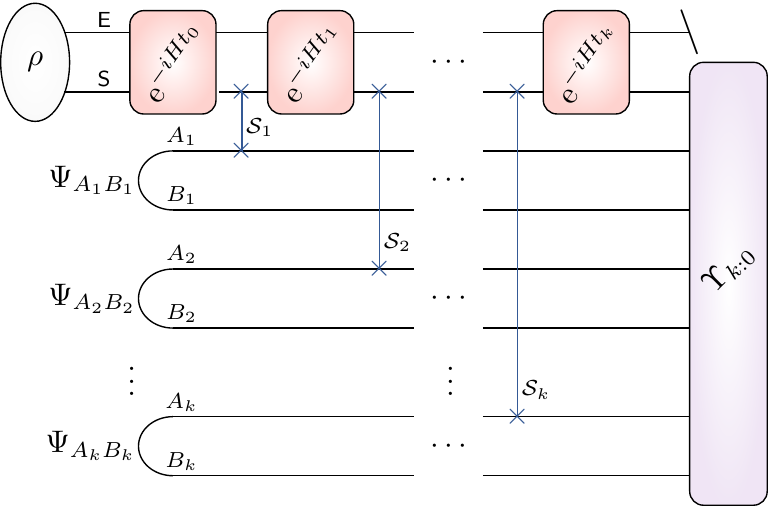}
\caption{\footnotesize{Circuit diagram of the Choi state of a process tensor corresponding to definition~\eqref{PT Choi state Def}.}}\label{PT circuit diagram}
\end{figure}

The associated Choi state of a time-evolved process tensor with initial state $\rho$ is then given by
\begin{gather}
	\Upsilon_{k:0}=\tr_E[\,\mc{U}_{k:0}(\rho\otimes\psi^{\otimes{k}})\,\mc{U}_{k:0}^\dg],
	\label{PT Choi state Def}
\end{gather}
where $\psi=\sum|ii\rangle\!\langle{jj}|$ is maximally entangled and unnormalized, and where here
\begin{gather}	\mc{U}_{k:0}:=(U_k\otimes\mbb1)\mc{S}_k\cdots(U_1\otimes\mbb1)\mc{S}_1(U_0\otimes\mbb1),\end{gather}
with all identity operators $\mbb1$ in the total ancillary system and with the $U_i$ being $\mathsf{SE}$ unitary operators at step $i$, and
\begin{gather}\mc{S}_i:=\sum_{\alpha,\beta}\mf{S}_{\alpha\beta}\otimes\mbb1_{_{A_1B_1\cdots{A}_{i-1}B_{i-1}}}\otimes|\beta\rangle\!\langle\alpha|\otimes\mbb1_{_{B_iA_{i+1}B_{i+1}\cdots{A}_kB_k}},\end{gather}
with $\mf{S}_{\alpha\beta}=\mbb1_E\otimes|\alpha\rangle\!\langle\beta|$. This can be visualised as the quantum circuit depicted in Figure~\ref{PT circuit diagram} when the unitary evolution is determined by a time-independent Hamiltonian, as we detail below, and we highlight that $\Upsilon$ is defined directly with the first input being the initial state $\rho$ (as opposed to half of a maximally entangled state in $\mathsf{S}$). Explicitly, it can be written as
\begin{align}\Upsilon_{k:0}=\sum\tr_E\left[{U}_k\FS_{\alpha_k\beta_k}
\cdots{U}_1\FS_{\alpha_1\beta_1}U_0\,\rho\,U_0^\dg
\FS_{\gamma_1\delta_1}^\dg{U}_1^\dg\cdots\FS_{\gamma_k\delta_k}^\dg{U}_k^\dg\right]\otimes|\beta_1\alpha_1\cdots\beta_k\alpha_k\rangle\!\langle\delta_1\gamma_1\cdots\delta_k\gamma_k|.\end{align}

 $\mf{S}_{\alpha\sigma}=\mf{S}_{\sigma\alpha}^\dg$, $\mf{S}_{ab}\mf{S}^\dg_{cd}=\delta_{bd}\mf{S}_{ac}$ and $\tr(\mf{S}_{ab})=d_E\delta_{ab}$. Also notice that the resulting Choi process tensor state belongs to the whole $\mathsf{S}$ plus ancillary system, which has dimension $d_S^{2k+1}$.

In particular, here we deal with evolution by time-independent Hamiltonians $H$, i.e., with $U_j=\exp[-iHt_j]$. Also, as done in~\cite{ShortSystemsAndSub, ShortFinite, PhysRevE.79.061103}, we consider first a pure initial state $\rho=|\phi\rangle\!\langle\phi|$ and then extend our results to mixed initial states by purification; this also allows to choose an energy eigenbasis $\{|n\rangle\}$ for $H$ such that the evolution of the initial state is the same as that given by a non-degenerate Hamiltonian $H^\prime=\sum_n{E}_n|n\rangle\!\langle{n}|$, i.e.,
\begin{gather}
	\rho(t_0)=U_{0}\rho\,U^\dg_{0}=\sum_{m,n}e^{-it_0(E_m-E_n)}\rho_{mn}|m\rangle\!\langle{n}|,
\end{gather}
where $\rho_{mn}=\langle{m}|\rho|n\rangle=\langle{m}|\phi\rangle\!\langle\phi|n\rangle$.

\section{Proof of the main result}\label{Appendix: bound}
We consider the multi-time expectation for CP (weighted) maps $\mc{A}_i(\cdot)=\sum{a}_i^\mu{A}_i^\mu(\cdot){A}_i^{\mu\,\dg}$ acting on a subsystem $\mathsf{S}$ of a joint $\mathsf{SE}$ system, together with an ancillary space $\mathsf{\Gamma}$, where the weights are $a_i^\mu\in\mathbb{R}$ and with $\sum{A}_i^{\mu\,\dg}{A}_i^\mu\leq\mbb1$. The full initial $\mathsf{SE\Gamma}$-state $\varrho$ will be given by $\rho\otimes\gamma$, where $\gamma$ acts on $\mathsf{\Gamma}$. This expectation in a process $\Upsilon$ is given by
\begin{equation}
    \langle\Lambda\rangle_\Upsilon=\tr[\mc{A}_k\,\mc{U}_k\cdots\mc{A}_0\,\mc{U}_0(\varrho)],
\end{equation}
where implicitly $\mc{A}_i$'s act only on $\mathsf{S\Gamma}$, while the unitaries $\mc{U}_i(\cdot)=U_i(\cdot)U_i^\dg$ with $U_\ell=\exp(-iH_\ell t)$ act on the $\mathsf{SE}$ system. We consider a fixed set of projectors $\{P_n\}$ for all Hamiltonians such that $H_\ell=\sum P_nE_{n_\ell}$ at each step $\ell$, with $P_n$ projecting onto the energy eigenspaces of $H_i$ with energy $E_{n_i}$. We also denote simply by $\cdot$ the composition of superoperators when clear by context. Let
\begin{equation}
    \varpi\equiv\varpi_0\equiv\lim_{T_0\to\infty}\int_0^\infty\mc{U}_0(\varrho)\,\mathscr{P}_{T_0}\,dt_0=\lim_{T_0\to\infty}\int_0^\infty\mc{U}_0(\rho)\,\mathscr{P}_{T_0}(t_0)\,dt_0\otimes\gamma,
\end{equation}
for a probability distribution on all $t_0>0$ with density function $\mathscr{P}_{T_0}$ with a parameter $T_0$, and similarly,
\begin{align}
    \varpi_1&\equiv\lim_{T_1\to\infty}\int_0^\infty\mc{U}_1\mc{A}_0(\varpi_0)\,\mathscr{P}_{T_1}(t_1)\,dt_1,\\
    &\vdots\nonumber\\
    \varpi_k&\equiv\lim_{T_k\to\infty}\int_0^\infty\mc{U}_k\mc{A}_{k-1}(\varpi_{k-1})\,\mathscr{P}_{T_k}\,dt_k(t_k),
\end{align}
for all $t_1,\ldots,t_k$. These probability functions are set to be such that
\begin{equation}
    \lim_{T_\ell\to\infty}\int_0^\infty\,\exp[-it(E_{n_\ell}-E_{m_\ell})]\,\mathscr{P}_{T_\ell}(t)\,dt=\delta_{n_\ell m_\ell},
\end{equation}
with the parameters $T_j$ taking the role of an uncertainty parameter, e.g. the variance of a given distribution.

We now define $\langle\Lambda\rangle_\Omega\equiv\lim_{\mathbf{T}\to\infty}\overline{\langle\Lambda\rangle_\Upsilon}^{\mathscr{P}_\mathbf{T}}=\tr[\mc{A}_k(\varpi_k)]$ where the overline means time-average over all $t_i$, i.e.
\begin{equation}
    \overline{X}^{\mathscr{P}_\mathbf{T}}=\int_0^\infty\mathscr{P}_{T_k}(t_k)\,dt_k\cdots\int_0^\infty\,\mathscr{P}_{T_0}(t_0)\,dt_0\,X(t_0,\ldots,t_k).
\end{equation}
In general, we can write
\begin{align}
    \varpi_i&=\sum{P}_n\mc{A}_{i-1}(P_{n^\prime}\mc{A}_{i-2}(\cdots\mc{A}_0(\varpi_0)\cdots{P}_{n^\prime})P_{n},
\end{align}
for any $0\leq{i}\leq{k}$. Let us denote by $\mc{P}_{nm}=P_n(\cdot)P_m$ and $\mc{D}(X)=\sum_{n}\mc{P}_{nn}(\cdot)=\sum_nP_n(\cdot)P_n$ the dephasing map with respect to $\{P_n\}$, then we can similarly write
\begin{align}
    \varpi_i&=\mc{D}\mc{A}_{i-1}\mc{D}\cdots\mc{A}_0\mc{D}(\varrho),
\end{align}
and so
\begin{equation}
    \langle\Lambda\rangle_\Omega=\tr[\mc{A}_k\mc{D}\mc{A}_{k-1}\mc{D}\cdots\mc{A}_0\mc{D}(\varrho)].
\end{equation}

We now consider the difference
\begin{align}
	|\langle\Lambda\rangle_{\overline{\Upsilon}^{\mathscr{P}_\mathbf{T}}-\Omega}|\equiv|\tr[(\overline{\Upsilon}^{\mathscr{P}_\mathbf{T}}-\Omega)\Lambda]|,
\end{align}
where
\begin{equation}
    \langle\Lambda\rangle_{\overline{\Upsilon}^{\mathscr{P}_\mathbf{T}}}=\sum_{n,m}\overline{\mathfrak{E}(\mathbf{t},\mathbf{n},\mathbf{m})}^{\mathscr{P}_\mathbf{T}}\tr[\mc{A}_k\mc{P}_{nm}\mc{A}_{k-1}\cdots\mc{A}_0\mc{P}_{nm}(\rho)],
\end{equation}
with
\begin{equation}
    \overline{\mathfrak{E}(\mathbf{t},\mathbf{n},\mathbf{m})}^{\mathscr{P}_\mathbf{T}}\equiv\overline{\exp[-it_k(E_{n_k}-E_{m_k})]|}^{\mathscr{P}_{T_k}}\cdots\overline{\exp[-it_0(E_{n_0}-E_{m_0})]}^{\mathscr{P}_{T_0}}.
\end{equation}

Defining $G_{n_\ell{m}_\ell}^{(\ell)}\equiv\overline{\exp[-it_\ell(E_{n_\ell}-E_{m_\ell})]|}^{\mathscr{P}_{T_\ell}}$, this is $\overline{\mathfrak{E}(\mathbf{t},\mathbf{n},\mathbf{m})}^{\mathscr{P}_\mathbf{T}}=\prod_{\ell=0}^kG_{n_\ell{m}_\ell}^{(\ell)}$. Let us define the partial dephasing superoperator $\mc{G}_\ell$ by $\mc{G}_\ell(\rho) = \sum_{n_\ell, m_\ell} G^{(\ell)}_{n_\ell, m_\ell} \mc{P}_{nm}(\rho)$ and $\mc{G}_{k:\ell}:= \bigcircop_{j=\ell}^k \mc{G}_j$, such that $\mc{G}_{k:\ell}(\rho) = \sum_{n,m} G^{(k)}_{n_k, m_k}\cdots{G}^{(\ell)}_{n_\ell, m_\ell}\mc{P}_{nm}(\rho)$. Whenever we denote $\mc{G}(\varrho)$, this means $\mc{G}(\rho)\otimes\gamma$, and similarly for any other maps.

Let us look first at the case $k=1$ (we label with a subindex the step at which $\mc{D}$ is applied where relevant),
\begin{align}
    \left|\langle\Lambda\rangle_{\overline{\Upsilon}^{\mathscr{P}_\mathbf{T}}-\Omega}\right|&=\left|\tr\left[\left(\bigcircop_{j=0}^1 \mathcal{A}_j  \mc{G}_j - \bigcircop_{j=0}^1 \mathcal{A}_j \mc{D}_j\right)(\varrho)\right]\right|\nonumber \\
    &= \Bigg|\tr\left[\mc{A}_1\mc{A}_0 \left(\mc{G}_1\mc{G}_0 - \mc{D}_1\mc{D}_0\right)(\varrho)\right] +\tr\Big[\mathcal{A}_1\Big([\mc{G}_1,\mc{A}_0]\mc{G}_0 - [\mc{D}_{1}, \mc{A}_0]\mc{D}_0\Big)(\varrho)\Big]\Bigg|\nonumber\\
    &\leq\left|\tr\left[\mc{A}_1\mc{A}_0 \left(\mc{G}_1\mc{G}_0 - \mc{D}\right)(\varrho)\right]\right| + \left|\tr\left\{\mathcal{A}_{1} [\mc{G}_{1} - \mc{D}, \mc{A}_0] \mc{G}_0(\varrho)\right\}\right| + \left|\tr\left\{\mathcal{A}_{1} [\mc{D},\mc{A}_0]  (\mc{G}_0-\mc{D})(\varrho)\right\}\right|,
\end{align}
where the third line follows by the triangle inequality ($|a-c|\leq|a-b|+|b-c|$, here with $b=\tr\{\mc{A}_1[\mc{D},\mc{A}_0]\mc{G}_0(\varrho)\}$). Similarly, for general $k$,
\begin{align}
    \left|\langle\Lambda\rangle_{\overline{\Upsilon}^{\mathscr{P}_\mathbf{T}}-\Omega}\right|&=\left|\tr\left[\left(\bigcircop_{j=0}^{k} \mathcal{A}_j \mc{G}_j - \bigcircop_{j=0}^{k} \mathcal{A}_j \mc{D}_j\right)(\varrho)\right]\right|\nonumber \\
    &= \Bigg|\tr\left[\mc{A}_{k:0} \left(\mc{G}_{k:0} - \mc{D}_{k:0}\right)(\varrho)\right] \nonumber \\
    &\qquad + \sum_{\ell=0}^{k-1}\tr\Big[\mathcal{A}_{k:\ell+1} \Big([\mc{G}_{k:\ell+1}, \mc{A}_\ell]  \mc{G}_\ell \bigcircop_{j=0}^{\ell-1} \mathcal{A}_j \mc{G}_j- [\mc{D}_{k:\ell+1}, \mc{A}_\ell]\mc{D}_\ell \bigcircop_{j=0}^{\ell-1} \mathcal{A}_j \mc{D}_j \Big)(\varrho)\Big]\Bigg|\nonumber\\
    \leq& \left|\tr\left[\mc{A}_{k:0} \left(\mc{G}_{k:0} - \mc{D}\right)(\varrho)\right]\right|\nonumber \\
    & + \sum_{\ell=0}^{k-1}\left|\tr\left[\mathcal{A}_{k:\ell+1} [\mc{G}_{k:\ell+1} - \mc{D}, \mc{A}_\ell] (\varrho_\ell)\right]\right| + \sum_{\ell=0}^{k-1}\left|\tr\left[\mathcal{A}_{k:\ell+1} [\mc{D}, \mc{A}_\ell]  (\varrho_\ell-\varpi_\ell)\right]\right|,
\end{align}
where $\varrho_\ell:=\mc{G}_\ell\bigcircop_{j=0}^{\ell-1}\mc{A}_j \mc{G}_j(\varrho)$. Using the Schwartz inequality as $|\tr \mc{X}(\varrho)|= |\langle\!\langle X | \varrho \rangle\!\rangle| \leq \|\mc{X}\|\|\varrho\|_2$, where here we denote $\|\mc{X}\|:= \sup_{\|\sigma\|_2=1} \|\mc{X}(\sigma)\|_2$ for simplicity, we further find
\begin{align}
\left|\langle\Lambda\rangle_{\overline{\Upsilon}^{\mathscr{P}_\mathbf{T}}-\Omega}\right|&\leq \left\|\mc{A}_{k:0}\right\|\, \left\| \left(\mc{G}_{k:0} - \mc{D}\right)(\varrho)\right\|_2 + \sum_{\ell=0}^{k-1} \left\|\mathcal{A}_{k:\ell+1} \right\|\, \left\| [\mc{G}_{k:\ell+1} - \mc{D}, \mc{A}_\ell] ( \varrho_\ell)\right\|_2 \nonumber \\
    & \qquad \qquad + \sum_{\ell=0}^{k-1}\left\|\mathcal{A}_{k:\ell+1} \right\|\, \left\|[\mc{D}, \mc{A}_\ell] ( \varrho_\ell-\varpi_\ell)\right\|_2\nonumber\\
    &\leq \left\|\mc{A}_{k:0}\right\|\, \left\| \left(\mc{G}_{k:0} - \mc{D}\right)(\varrho)\right\|_2 + \sum_{\ell=0}^{k-1} \left\|\mathcal{A}_{k:\ell+1} \right\|\,\left\| \mc{A}_\ell\right\|\, \left\|(\mc{G}_{k:\ell+1} - \mc{D}) ( \varrho_\ell)\right\|_2 \nonumber \\
    & \qquad \qquad + \sum_{\ell=0}^{k-1} \left\|\mathcal{A}_{k:\ell+1} \right\|\, \left\| (\mc{G}_{k:\ell+1} - \mc{D})\mc{A}_\ell ( \varrho_\ell)\right\|_2 + \sum_{\ell=0}^{k-1}\left\|\mathcal{A}_{k:\ell+1} \right\|\, \left\|[\mc{D}, \mc{A}_\ell] ( \varrho_\ell-\varpi_\ell)\right\|_2.
    \label{appendix eq: 2normbound}
\end{align}

For the first term of Eq.~\eqref{appendix eq: 2normbound}, we have $\mc{G}_{k:\ell} - \mc{D} = \sum_{n \neq m} G_{n_\ell{m}_\ell}^{(\ell)}\cdots{G}_{n_k{m}_k}^{(k)}\mc{P}_{nm}$, therefore,
\begin{align}
    \left\| \left(\mc{G}_{k:0} - \mc{D}\right)(\varrho)\right\|_2^2&=\tr\left|\sum_{n \neq m} G_{n_k{m}_k}^{(k)}\cdots{G}_{n_0m_0}^{(0)} \mc{P}_{nm}(\varrho)\right|^2\nonumber\\
    &=\sum_{\substack{n \neq m \\ n^\prime \neq m^\prime}}\prod_{j=0}^k G_{n_jm_j}^{(j)}G_{m_j^\prime{n}_j^\prime}^{(j)}\tr\left[ \mc{P}_{nm}(\varrho)\mc{P}_{m^\prime{n}^\prime}(\varrho)\right]\nonumber\\
    &=\sum_{n \neq m }\prod_{j=0}^k |G_{n_jm_j}^{(j)}|^2\tr\left[ P_n \varrho P_m\varrho\right]\nonumber\\
    &\leq\prod_{j=0}^k\max_{n\neq{m}}|G_{n_jm_j}^{(j)}|^2\left\{\sum_{n , m}\tr[P_n\varrho P_m \varrho]-\sum_n\tr[P_n\varrho P_n \varrho]\right\}\nonumber\\
    &=\prod_{j=0}^k\max_{n\neq{m}}|G_{n_jm_j}^{(j)}|^2\tr(\varrho^2-\varpi^2)\nonumber\\
    &=\|\varrho-\varpi\|_2^2\prod_{j=0}^k\max_{n\neq{m}}|G_{n_jm_j}^{(j)}|^2.
\end{align}
where we used $\tr(\varrho^2-\varpi^2)=\|\varrho-\varpi\|_2^2$, because $\tr(\varrho\,\varpi)=\tr(\varpi^2)$. This, together with Eq.~\eqref{appendix eq: 2normbound} already gives the result in Eq.~\eqref{eq: result main} for the case $H_i=T_j$ and $T_i=T_j$ for all $i\neq{j}$.

For the second term then, similarly,
\begin{align}
    \left\| \left(\mc{G}_{k:\ell+1} - \mc{D}\right)(\varrho_\ell)\right\|_2^2
    &\leq\prod_{j=\ell+1}^k\max_{n\neq{m}}|G_{n_jm_j}^{(j)}|^2\left\{\sum_{n , m}\tr[P_n\varrho_\ell P_m \varrho_\ell]-\sum_n\tr[P_n\varrho_\ell P_n \varrho_\ell]\right\}\nonumber\\
    &=\prod_{j=\ell+1}^k\max_{n\neq{m}}|G_{n_jm_j}^{(j)}|^2\tr[\varrho_\ell^2-\mc{D}(\varrho_\ell)\varrho_\ell]\nonumber\\
    &=\|\varrho_\ell-\mc{D}(\varrho_{\ell})\|_2^2\prod_{j=\ell+1}^k\max_{n\neq{m}}|G_{n_jm_j}^{(j)}|^2,
\end{align}
as $\tr[(\mc{D}(\varrho_\ell))^2]=\tr[\mc{D}(\varrho_\ell)\varrho_\ell]$.

For the third term, with $\varrho^\prime_\ell=\mc{A}_\ell(\varrho_\ell)$,
\begin{align}
    \left\| \left(\mc{G}_{k:\ell+1} - \mc{D}\right)(\varrho_\ell^\prime)\right\|_2^2
    &\leq\prod_{j=\ell+1}^k\max_{n\neq{m}}|G_{n_jm_j}^{(j)}|^2\left\{\sum_{n , m}\tr[P_n(\varrho_\ell^\prime) P_m (\varrho_\ell^\prime)]-\sum_n\tr[P_n(\varrho_\ell^\prime) P_n (\varrho_\ell^\prime)]\right\}\nonumber\\
    &=\prod_{j=\ell+1}^k\max_{n\neq{m}}|G_{n_jm_j}^{(j)}|^2\tr[\varrho_\ell^{\prime\,2}-\mc{D}(\varrho_\ell^\prime)\varrho_\ell^\prime]\nonumber\\
    &=\|\mc{A}_\ell(\varrho_\ell)-\mc{D}\mc{A}_\ell(\varrho_\ell)\|_2^2\prod_{j=\ell+1}^k\max_{n\neq{m}}|G_{n_jm_j}^{(j)}|^2.
\end{align}

Finally, for the fourth term, as
\begin{align}
    \|[\mc{D},\mc{A}_\ell](\varrho_\ell-\varpi_\ell)\|_2&\leq\|\mc{D}(\varrho_{\ell+1})-\varpi_{\ell+1}\|_2+\|\mc{A}_\ell\|\|\mc{D}(\varrho_\ell)-\varpi_\ell\|_2,
\end{align}
then, let us denote $\mf{S}_{b:a}:=\prod_{j=a}^b\max_{n\neq{m}}|G_{n_jm_j}^{(j)}|$, so that putting all together,
\begin{align}
    \left|\langle\Lambda\rangle_{\overline{\Upsilon}^{\mathscr{P}_\mathbf{T}}-\Omega}\right|&\leq \mf{S}_{k:0}\,\|\mc{A}_{k:0}\|\,\|\varrho-\varpi\|_2\nonumber\\
    &\quad +\sum_{\ell=0}^{k-1}\mf{S}_{k:\ell+1}\,\|\mc{A}_{k:\ell+1}\|\,\bigg\{\|\mc{A}_\ell\|\|\varrho_\ell-\mc{D}(\varrho_{\ell})\|_2+\|\mc{A}_\ell(\varrho_\ell)-\mc{D}\mc{A}_\ell(\varrho_\ell)\|_2\bigg\}\nonumber\\
    &\quad\quad +\sum_{\ell=0}^{k-1}\|\mc{A}_{k:\ell+1}\|\left\{\|\mc{D}(\varrho_{\ell+1})-\varpi_{\ell+1}\|_2+\|\mc{A}_\ell\|\|\mc{D}(\varrho_\ell)-\varpi_\ell\|_2\right\},
\end{align}
as discussed in the main text.

\end{document}